\documentclass[8.5pt,twoside]{article}

\usepackage[super,sort&compress,comma]{natbib}
\usepackage{times,mathptm}
\usepackage{sectsty}
\usepackage{balance}
\usepackage[text={11.3cm,20.4cm},centering]{geometry}

\usepackage{graphicx} 
\usepackage{lastpage}
\usepackage[format=plain,justification=raggedright,singlelinecheck=false,font=small,labelfont=bf,labelsep=space]{caption}
\usepackage{fancyhdr}
\usepackage{amsmath}
\usepackage{latexsym}
\usepackage{amsfonts}
\usepackage[psamsfonts]{amssymb}
\usepackage[mathlines]{lineno}

\usepackage{url}

\usepackage[
hyperindex=true,
pdftitle={},
colorlinks=true,
pagebackref=false,
citecolor=blue,
plainpages=false,
pdfpagelabels,
breaklinks=true,
linkcolor=blue,
urlcolor=blue,
 runcolor=blue,
 anchorcolor=blue
]{hyperref}
\usepackage{microtype}

\renewcommand{\epsilon}{\text{\usefont{OML}{cmr}{m}{n}\symbol{15}}}

\begin{document}

\thispagestyle{plain}
\fancypagestyle{plain}{
\fancyhead[L]{\includegraphics[height=8pt]{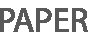}}
\fancyhead[C]{\hspace{-1cm}\includegraphics[height=15pt]{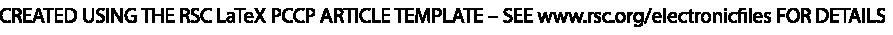}}
\fancyhead[R]{\includegraphics[height=10pt]{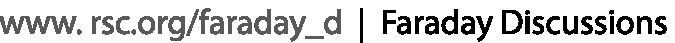}\vspace{-0.2cm}}
\renewcommand{\headrulewidth}{1pt}}
\renewcommand{\thefootnote}{\fnsymbol{footnote}}
\renewcommand\footnoterule{\vspace*{1pt}%
\hrule width 11.3cm height 0.4pt \vspace*{5pt}}
\setcounter{secnumdepth}{5}

\makeatletter
\renewcommand{\fnum@figure}{\textbf{Fig.~\thefigure~~}}
\def\subsubsection{\@startsection{subsubsection}{3}{10pt}{-1.25ex plus -1ex minus -.1ex}{0ex plus 0ex}{\normalsize\bf}}
\def\paragraph{\@startsection{paragraph}{4}{10pt}{-1.25ex plus -1ex minus -.1ex}{0ex plus 0ex}{\normalsize\textit}}
\renewcommand\@biblabel[1]{#1}
\renewcommand\@makefntext[1]%
{\noindent\makebox[0pt][r]{\@thefnmark\,}#1}
\makeatother
\sectionfont{\large}
\subsectionfont{\normalsize}

\fancyfoot{}
\fancyfoot[LO,RE]{\vspace{-7pt}\includegraphics[height=8pt]{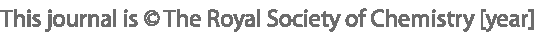}}
\fancyfoot[CO]{\vspace{-7pt}\hspace{5.9cm}\includegraphics[height=7pt]{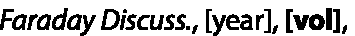}}
\fancyfoot[CE]{\vspace{-6.6pt}\hspace{-7.2cm}\includegraphics[height=7pt]{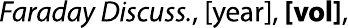}}
\fancyfoot[RO]{\scriptsize{\sffamily{1--\pageref{LastPage} ~\textbar  \hspace{2pt}\thepage}}}
\fancyfoot[LE]{\scriptsize{\sffamily{\thepage~\textbar\hspace{3.3cm} 1--\pageref{LastPage}}}}
\fancyhead{}
\renewcommand{\headrulewidth}{1pt}
\renewcommand{\footrulewidth}{1pt}
\setlength{\arrayrulewidth}{1pt}
\setlength{\columnsep}{6.5mm}
\setlength\bibsep{1pt}

\noindent\LARGE{\textbf{Coherent Phase Control in Closed and Open Quantum Systems: A General Master Equation Approach
}
\vspace{0.6cm}

\noindent\large{\textbf{Leonardo A. Pach\'on\textit{$^{a,b}$},
Li Yu\textit{$^{b,c}$} and
Paul Brumer$^{\ast}$\textit{$^{b}$}}}\vspace{0.5cm}

\noindent\textit{\small{\textbf{Received Xth XXXXXXXXXX 2012, Accepted Xth XXXXXXXXX 20XX\newline
First published on the web Xth XXXXXXXXXX 200X}}}

\noindent \textbf{\small{DOI: 10.1039/c000000x}}
\vspace{0.6cm}

\noindent \normalsize{The underlying mechanisms for one photon phase control are revealed 
through a master equation approach.  Specifically, two mechanisms are identified, 
one operating on the laser time scale 
and the other on the time scale of the system-bath interaction.  The effect of the 
secular and non-secular Markovian 
approximations are carefully examined. }
\vspace{0.5cm}

\section{Introduction:}


\footnotetext{\textit{$^{a}$~Instituto de F\'isica,
Universidad de Antioquia, AA 1226, Medell\'in, Colombia.
Tel: 57 4 219 6436;
E-mail: leonardo.pachon@fisica.udea.edu.co}}
\footnotetext{\textit{$^{b}$~Chemical Physics Theory Group,
Department of Chemistry and
Center for Quantum Information and Quantum Control,
University of Toronto, Toronto, Canada M5S 3H6.
Fax: 1 416 978 5325; Tel: 1 416 978 7044;
E-mail: pbrumer@chem.utoronto.ca}}
\footnotetext{\textit{$^{c}$~Department of Physics,
Harvard University, Cambridge, Massachusetts 02138, USA;
E-mail: lyu@fas.harvard.edu }}


The possibility of manipulating and controlling quantum systems by using quantum features
is the goal of many modern control protocols\cite{SB11}.
An unusual type of control, one photon phase control (OPPC), has been the subject of recent attention
and discussion (see, e.g., Faraday Discussions 153 \cite{faraday153}).
In this case, control takes place by varying the relative {\it phase} of components of a weak
pulse, while keeping the power spectrum of the light source fixed.
A seminal proof\cite{BS89} showed that one-photon phase control was not possible for isolated
molecular systems in which control is over products in the continuum.
A subsequent experiment on control of retinal isomerization in bacteriorhodopsin\cite{PM06}
motivated controversy\cite{J07,PM07,bucksbaum,millerreply2}
and the need for clarification of conditions under which such control was possible.
This clarification, provided in Ref.~\citenum{SAB10} showed that control was {\it possible} for both
closed systems and for open quantum systems under well defined conditions.
Specifically, we showed the following for control of an observable $\hat{O}$ in a system defined
by Hamiltonian $\hat{H}_{\mathrm M}$:

(a)  If the system is closed (i.e., not coupled to an environment), then one-photon phase control
is possible if $[\hat{H}_{\mathrm M},\hat{O}] \ne 0$.
For example, control over isomerization in an isolated molecule is possible since the probability
of observing an isomer (e.g., \textit{cis} or \textit{trans}) is an observable that does not commute
with $\hat{H}_{\mathrm M}$.
The situation is similar for control over product formation in radiationless transitions, such as
intersystem crossing and internal conversion\cite{GSB13}.

However, if $[\hat{H}_{\mathrm M},\hat{O}] = 0$, then one-photon phase control is not possible in an isolated
system.

(b)  If the system is coupled to an environment then, as above, control is possible if
$[\hat{H}_{\mathrm M},\hat{O}] \ne 0$.
However, even in the case where $[\hat{H}_{\mathrm M},\hat{O}] = 0$ control may still be possible, in which
case it is \textit{environmentally assisted}.

Reference \citenum{SAB10} provided these general rules, derived in a somewhat different way
below, but did not provide any insight into the way in which environmental assistance works,
or the conditions under which such control is appreciable.
This is the subject of this paper.
Specifically, we demonstrate the significant role of non-Markovian dynamics and strong
system-bath coupling in enhancing the extent of one-photon phase control.
To do so we utilize a master equation based approach to the open system case, which,
as seen below, proves enlightening.
The net result of this paper is the demonstration of the underlying mechanism responsible
for environmentally assisted one-photon phase control.

Note, also, as an aside, that we have also computationally demonstrated \cite{AB13}
one-photon phase control in a model retinal system, making a connection with the original
experiments on related systems.

\section{One-photon phase control and closed quantum systems}

Let us consider a molecular system M described by the Hamiltonian $\hat{H}_{\mathrm{M}}$
with $\hat{H}_{\mathrm{M}} |a\rangle = E_a | a\rangle$.
If the system state is described by the density operator $\hat{\rho}_{\mathrm{M}} $,
the expectation value of a given observable $\hat{O}$ of system M at time $t$ can be calculated as
\begin{equation}
\langle \hat{O}(t) \rangle = \sum_{a,b}\langle a | \hat{O}  | b \rangle
\langle b | \hat{\rho}_{\mathrm{M}}  | a \rangle.
\end{equation}
If $[\hat{O},\hat{H}_{\mathrm{M}}] = 0$, we simply get
\begin{equation}
\label{equ:ExpValOcomH}
\langle \hat{O}(t) \rangle = \sum_{a}\langle a | \hat{O}  | a \rangle
\langle a | \hat{\rho}_{\mathrm{M}}  | a \rangle.
\end{equation}
For this case, only diagonal terms contribute to the time evolution of the expectation
value of the observable $\hat{O}$.

The essence of phase control is trying to manipulate the expectation of $\hat{O}$ by changing
the \textit{phase} of an applied laser pulse.
When the laser pulse is taking into account, the Hamiltonian can be re-written as
$\hat{H}_0 = \hat{H}_{\mathrm{M}} + \hat{H}^{\mathrm{MR}} =  \hat{H}_{\mathrm{M}}
- \hat{\mathbf{d}} \cdot \mathbf{E}(z,t)$, with
$\mathbf{E}(z,t) = \boldsymbol{\varepsilon} \mathcal{E}(z,t)$ where
\begin{equation}
\mathcal{E}(z,t) = \int_{-\infty}^{\infty} \mathrm{d} \omega\, \epsilon(\omega)
\exp\left[-\mathrm{i} \omega\left(t-\frac{z}{c}\right)\right],
\label{equ:ElectFieldStrenght}
\end{equation}
is the amplitude of the field with polarization vector $ \boldsymbol{\varepsilon}$,
which can be expressed in terms of frequency spectrum
$\epsilon(\omega)= |\epsilon(\omega)|\exp[\mathrm{i}\phi(\omega)]$ and, $\hat{\mathbf{d}}$ is the transition-dipole operator.
We are interested in manipulating the time evolution of the observable $\hat{O}$
by means of varying the spectral phase $\phi(\omega)$ only.

In order to clearly understand the role of the environment in  OPPC,
it is illustrative to analyze in detail the circumstances under which\cite{SAB10}
phase control is not possible.
The evolution of  the density matrix, for unitary time-evolution, is governed by
\begin{equation}
\label{equ:UniEvoGenCase}
\dot{\rho}_{ab} = - \mathrm{i} \omega_{ab} \rho_{ab}
+ \frac{\mathrm{i}}{\hbar} \sum_c \left[H^{\mathrm{MR}}_{ac}(t)\rho_{cb} -
\rho_{ac} H^{\mathrm{MR}}_{cb}(t)\right]
\end{equation}
where $\rho_{ab}  = \langle a | \hat{\rho}_{\mathrm{M}} | b \rangle$, $\hbar \omega_{ab} = E_a
- E_b$ and $H^{\mathrm{MR}}_{ab}(t) =  \mathrm{d}_{ab} \mathcal{E}(z,t)
= \langle a |\boldsymbol{\varepsilon}\cdot \hat{\mathbf{d}}\, | b\rangle \mathcal{E}(z,t)$.
\emph{Due the unitary character of the dynamics}, we can integrate
Eq.~(\ref{equ:UniEvoGenCase}) by means of the following ansatz for the
wavefunction
\begin{equation}
\label{equ:wfunctansatz}
|\Psi \rangle = \sum_n b_n(t) |E_n\rangle \exp\left(-\frac{\mathrm{i}}{\hbar}E_n t \right),
\end{equation}
with $\hat{\rho}_{\mathrm{M}} (t) = |\Psi (t)\rangle \langle \Psi(t) |$.

Let us consider, e.g., that the molecule is initially ($t=-\infty$) in the pure state $|E_1\rangle$,
so we have $b_1(t=-\infty)=1$ and $b_k(t=-\infty) = 0$ for $k\neq1$.
After formally integrating Eq.~(\ref{equ:wfunctansatz}) for $b_m$, we get
\begin{equation}
b_m(t) = \int_{-\infty}^{t} \mathrm{d}s\,b_1(s) \exp(\mathrm{i}\omega_{m,1}s)
H^{\mathrm{MR}}_{m1}(s).
\end{equation}
\emph{If the perturbation is weak, for unitary evolution}, one can assume that $b_1(t)=1$ at all
times, thus
\begin{equation}
b_m(t) = \frac{\mathrm{i}}{\hbar}\mathrm{d}_{m1}
\int_{-\infty}^{\infty}\mathrm{d}\omega \bar{\epsilon}(\omega)
\int_{-\infty}^{t} \mathrm{d}s\, \exp[\mathrm{i}(\omega_{m,1}-\omega)s],
\end{equation}
where we have made use of Eq.~(\ref{equ:ElectFieldStrenght}) and defined
$\bar{\epsilon}(\omega) = \epsilon(\omega) \exp(\mathrm{i} \omega z/c)$.
Since our interest is in the long time regime, we extend the upper integration limit
to infinity and obtain $2\pi \delta(\omega_{m1} - \omega)$ for the integral over $s$.
Finally, we have
\begin{equation}
b_m(\infty) = 2\pi \frac{\mathrm{i}}{\hbar}\mathrm{d}_{m1}|\epsilon(\omega_{m1})|
\exp\left[\mathrm{i}\left(\phi(\omega_{m1}) + \frac{1}{c}\omega_{m1}z \right) \right],
\end{equation}
which yields
\begin{align}
\label{equ:DentMatElemUnitEvo}
\begin{split}
\rho_{ab}(t) &= \frac{4 \pi}{\hbar^2}
\mathrm{d}_{a1} \mathrm{d}^{*}_{b1}|\epsilon(\omega_{a1})| |\epsilon(\omega_{b1})|
\exp(-\mathrm{i}\omega_{ab}t)
\\ &\times
\exp\left\{\mathrm{i}\left[\phi(\omega_{a1}) -\phi(\omega_{b1})
+ \frac{z}{c}(\omega_{a1} - \omega_{b1}) \right] \right\}.
\end{split}
\end{align}

From here, it is clear that no spectral phase information is present in the diagonal
terms, $a=b$, and therefore it is impossible to manipulate the time evolution of the observable
$\hat{O}$ in Eq.~(\ref{equ:ExpValOcomH}).

Based on the equation (\ref{equ:DentMatElemUnitEvo}), it is easy to track the vanishing of
the spectral phase contribution to the fact that $\hat{\rho}_{\mathrm{M}}$ was constructed as the projector
onto the state $|\Psi(t)\rangle$, i.e., to the fact that the evolution is unitary.
This can be understood as follows. Under unitary time evolution, the dynamics is generated
by the unitary time-evolution operator
$\hat{U}(t) = \hat{\mathcal{T}} \exp(-\mathrm{i}\int_{t_0}^t \mathrm{d}s \hat{H}(s)/\hbar)$,
by means of the relation $\hat{\rho}_{\mathrm{M}}(t) = \hat{U}(t) \hat{\rho}_{\mathrm{M}}(0) \hat{U}^{\dag}(t)$.
If $\hat{\rho}_{\mathrm{M}}(0)$ represents the density operator of a pure state $|\Psi\rangle$, then $\hat{U}(t)$
evolves $|\Psi\rangle$ (in the direct Hilbert space) whereas $\hat{U}^{\dag}(t)$ evolves $\langle\Psi|$
(in the dual Hilbert space).
Since the generator in the dual space, $\hat{U}^{\dag}(t)$, is just the time-reversed version
of the generator in the direct space, $\hat{U}(t)$, then the evolution in the two spaces interfere
in the density operator in such a way that no phase information is encoded in the diagonal
elements [see Eq.~(\ref{equ:DentMatElemUnitEvo})].
This will no longer be true in the case of non-unitary dynamics (e.g., for open quantum systems) 
because of time-symmetry breaking.
This fact can be identified as a first contribution from the environment to  OPPC (see below).

In what follows, we explore in detail the possible ways in which the environment can
support the generation of OPPC.

\section{One-photon phase control and open quantum systems.}%
The study of open-quantum-systems dynamics is a demanding and intricate task due
to the various time/energy and coupling strength scales involved.
In the case of molecular systems, it is common that one component of the 
system plays the
role of the environment.  For example, if one is interested in the electronic dynamics, then the
vibrational and rotational modes of the same system comprise the environment.
In this situation, one expects that the coupling to the environment is strong and the dynamics
are non-local in time, i.e., it depends on the previous conformational states of the molecular
system (non-Markovian dynamics)\cite{PB12,PB11}.
This non-local character of the dynamics has been proved to be vital in physical systems
under the influence of external fields\cite{PB12b,PB12c}
Additionally, there is no apparent justification for neglecting the initial correlations between the part
we name the system and the part we name the environment.
All these features prevent us from applying standard approximation schemes and somehow
hide the underlying physical processes.

In order to identify the contribution from the environment to the one-photon phase control,
we consider the following situation:
a molecular system M is at equilibrium with its environment, a stationary/incoherent
situation that is subsequently irradiated by a coherent laser pulse.
In the first part of our analysis, Sec.~(\ref{subsect:InfTNonUnitDynam}), we neglect the initial
correlation between the system and the environment and discuss how the non-unitary dynamics
induced by the environment could assist OPPC.
In the second part, Sec.~(\ref{subsect:InfInitSysBathCorrel}), we include the initial correlations
with the environment and show how they could enhance the role of the environment in
assisting OPPC.

\subsection{Influence of the non-unitary dynamics}
\label{subsect:InfTNonUnitDynam}
In order to consider the influence of the environment on the system in a consistent way,
we start by considering the total dynamics of the two parts by means of the total Hamiltonian
\begin{equation}
\label{equ:TotHam}
\hat{H} =
\hat{H}_{\mathrm{M}} + \hat{H}^{\mathrm{ME}} + \hat{H}_{\mathrm{E}} + \hat{H}^{\mathrm{MR}} ,
\end{equation}
where $ \hat{H}^{\mathrm{ME}}$ denotes the interaction between the system and
the environment and $\hat{H}_{\mathrm{E}}$ characterizes the environment.
Let us consider that the interaction between the system and the bath is of the general
form
\begin{equation}
\label{equ:HME}
\hat{H}^{\mathrm{ME}} = \sum_{u} \hat{K}^{u} \otimes \hat{\Phi}^{u},
\end{equation}
where $\hat{K}^u$ and $\hat{\Phi}^u$ are operators of the system and the environment,
respectively.
Note that $\hat{K}^u$ and $\hat{\Phi}^u$ could be non-Hermitian as long as
$\hat{H}^{\mathrm{ME}}$ is Hermitian.
At this point, we assume that the initial density operator $\hat{\rho}_{\mathrm{ME}}$ of the 
system$+$bath factorizes at $t=t_0$ into a product of the system part $\hat{\rho}_{\mathrm{M}}$ 
and the environment part $\hat{\rho}_{\mathrm{E}}$, that is as 
$\hat{\rho}_{\mathrm{ME}}(t_0) = \hat{\rho}_{\mathrm{M}}(t_0) \otimes \hat{\rho}_{\mathrm{E}}(t_0)$.

For this case, the non-Markovian evolution of the system density matrix elements can be cast in the
form (cf. Chapter 3 in Ref.~\citenum{MK01})
\begin{equation}
\label{equ:GenNonMarEvo}
\begin{split}
\dot{\rho}_{ab} &= - \mathrm{i} \omega_{ab} \rho_{ab}
+ \frac{\mathrm{i}}{\hbar} \sum_c \left(H^{\mathrm{MR}}_{ac}\rho_{cb} -
\rho_{ac} H^{\mathrm{MR}}_{cb}\right)
\\
&- \sum_{c,d}\int\limits_0^{t-t_0} \mathrm{d}\tau \left\{
\mathsf{M}_{cd,db}(-\tau)\mathrm{e}^{\mathrm{i}\omega_{da}\tau}
\rho_{ac}(t-\tau)
+\mathsf{M}_{ac,cd}(\tau) \mathrm{e}^{\mathrm{i}\omega_{bc}\tau}
\rho_{db}(t-\tau)\right.
\\
&\left. -\left[
\mathsf{M}_{db,ac}(-\tau) \mathrm{e}^{\mathrm{i}\omega_{bc}\tau}
+
\mathsf{M}_{db,ac}(\tau) \mathrm{e}^{\mathrm{i}\omega_{da}\tau}
\right]\rho_{cd}(t-\tau)\right\},
\end{split}
\end{equation}
where $\mathsf{M}_{ab,cd}(t)$ determines the time span for correlations and are called the
memory matrix elements (cf. Chapter 3 in Ref.~\citenum{MK01}).
They are defined by
\begin{equation}
\mathsf{M}_{ab,cd}(t) = \sum_{u,v}C_{uv}(t) K_{ab}^{u} K_{cd}^{v},
\end{equation}
with $K_{ab}^{u} =  \langle a | \hat{K}^u | b \rangle$, and satisfy the relation 
$\mathsf{M}^*_{ab,cd}(t) = \mathsf{M}_{dc,ba}(-t)$.
The reservoir correlation function $C_{uv}(t)$ is given by
\begin{equation}
\label{equ:Cunu}
C_{uv}(t) = \frac{1}{\hbar^2}\langle \Phi^u(t) \Phi^v(0)\rangle_{\mathrm{R}}
- \frac{1}{\hbar^2}\langle \Phi^u\rangle_{\mathrm{R}} \langle \Phi^v\rangle_{\mathrm{R}},
\end{equation}
where $\langle \Phi^u(t) \Phi^v(0)\rangle_{\mathrm{R}} =
\mathrm{tr}[\hat{\rho}_{\mathrm{E,equ}} \Phi^u(t) \Phi^v(0)]$.
The first line in Eq.~(\ref{equ:GenNonMarEvo}) describes the unitary evolution and is
completely equivalent to Eq.~(\ref{equ:UniEvoGenCase}).
Hence, if no phase control can be achieved from Eq.~(\ref{equ:UniEvoGenCase}), no phase
control can be achieved from this part of the evolution in Eq.~(\ref{equ:GenNonMarEvo}).
The remaining lines describe the non-unitary non-Markovian dynamics, characterized by
the non-local evolution implied by the integration from $0$ to $t-t_0$.
The most relevant feature of this second part of Eq.~(\ref{equ:GenNonMarEvo}) is the
fact that the off-diagonal elements affect the dynamics of the diagonal terms,
despite the weak field condition.

Consider then the case where  the system is initially in equilibrium with the bath and the
excitation follows.
For convenience, let us  write Eq.~(\ref{equ:GenNonMarEvo}) as
$\dot{\hat{\rho}}_{\mathrm{M}} = -\mathrm{i} \hat{\mathcal {L}}_{\mathrm{M}} \hat{\rho}_{\mathrm{M}} 
- \hat{\mathcal {D}} \hat{\rho}_{\mathrm{M}}
- \mathrm{i} \hat{\mathcal {L}}_{\mathrm{laser}}\hat{\rho}_{\mathrm{M}}$,
where we have introduced the superoperators 
$\hat{\mathcal {L}}_{\mathrm{M}} \,\cdot = [\hat{H}_{\mathrm{M}}, \cdot]/\hbar$,
$\hat{\mathcal {L}}_{\mathrm{laser}} \,\cdot = [\hat{H}_{\mathrm{laser}}, \cdot]/\hbar$ 
and the so-called dissipative or relaxation superoperator $\hat{\mathcal {D}}$, which cannot
be given via a Liouville superoperator as an abbreviation of a simple commutator.
In the absence of the laser field, the solution to Eq.~(\ref{equ:GenNonMarEvo}) 
can then be written as 
$\hat{\rho}_{\mathrm{M}}(t) = \hat{\mathcal{T}}
\exp[-\mathrm{i} \int_{t_0}^{t} \mathrm{d}s(\hat{\mathcal {L}}_{\mathrm{M}}(s) -
\mathrm{i}\hat{\mathcal {D}}(s))]\hat{\rho}_{\mathrm{M}}(t_0)$.
Thus, in this representation, our initial state is an eigenstate of the superoperator
$\hat{\mathcal{T}}\exp[-\mathrm{i} \int_{t_0}^{t} \mathrm{d}s(\hat{\mathcal {L}}(s) -
\mathrm{i}\hat{\mathcal {D}}(s))]$,
i.e.,
$\hat{\mathcal{T}}\exp[-\mathrm{i} \int_{t_0}^{t} \mathrm{d}s 
(\hat{\mathcal {L}}_{\mathrm{M}}(s) - \mathrm{i}\hat{\mathcal {D}}(s))]
\hat{\rho}_{\mathrm{M}} = \lambda_{\hat{\rho}_{\mathrm{M}}} \hat{\rho}_{\mathrm{M}}$.
For high enough temperature and weak coupling to the bath [cf. Ref.~\citenum{PT12} and reference therein],
one can show that in the long time regime, and in the absence of the laser, the density operator of the system 
is well characterized by the canonical distribution $\hat{\rho}_{\mathrm{M,can}}$, i.e.,
\begin{equation}
\label{equ:IncIniSta}
\hat{\rho}_{\mathrm{M,can}} =
\lim_{t\rightarrow \infty}\hat{\rho}_{\mathrm{M}}(t) = \frac{1}{Z}\mathrm{e}^{- \hat{H}_{\mathrm{M}}\beta}
=\frac{1}{Z} \sum_a \mathrm{e}^{- E_a \beta} |a \rangle \langle a |,
\end{equation}
where $\{|a \rangle\}$ are eigenstates of $\hat{H}_{{\mathrm{M}}}$, 
$Z=\mathrm{tr}(\mathrm{e}^{- \hat{H}_{\mathrm{M}}\beta})$ is the partition function
and $\beta=1/k_{\mathrm{B}}T$.
Thus, $\hat{\rho}_{\mathrm{M,can}}$ is an eigenstate of non-driven dynamics 
with eigenvalue 1, i.e.,
$\exp[-\mathrm{i} \int_{t_0}^{t} \mathrm{d}s(\hat{\mathcal {L}}_{\mathrm{M}}(s) - \mathrm{i}\hat{\mathcal {D}}(s))]
\hat{\rho}_{\mathrm{M,can}}  =  \hat{\rho}_{\mathrm{M,can}}$ or equivalently,
$ \hat{\rho}_{\mathrm{M,can}}$ is a zero mode of the generator
$\hat{\mathcal {L}}_{\mathrm{M}}(s) - \mathrm{i}\hat{\mathcal {D}}(s)$.
Thus, in this description the laser excitation finds  the system in an incoherent superposition
of energy eigenstates.

For this initial state, under the weak field assumption, we can solve Eq.~(\ref{equ:GenNonMarEvo})
perturbatively in the laser term,
\begin{align}
\dot{\hat{\rho}}_{\mathrm{M}}^{(0)}(t) &=
-\mathrm{i}(\hat{\mathcal {L}}_{\mathrm{M}} - \mathrm{i}\hat{\mathcal {D}})
\hat{\rho}_{\mathrm{M,can}},
\\
\dot{\hat{\rho}}_{\mathrm{M}}^{(1)}(t) &=
- \mathrm{i} (\hat{\mathcal {L}}_{\mathrm{M}} - \mathrm{i}\hat{\mathcal {D}}) \hat{\rho}_{\mathrm{M}}^{(0)}(t)
 - \mathrm{i} \hat{\mathcal {L}}_{\mathrm{laser}}\hat{\rho}_{\mathrm{M}}^{(0)}(t) ,
\\
\begin{split}
\dot{\hat{\rho}}_{\mathrm{M}}^{(2)}(t) &=
- \mathrm{i} (\hat{\mathcal {L}}_{\mathrm{M}} - \mathrm{i}\hat{\mathcal {D}}) \hat{\rho}_{\mathrm{M}}^{(1)}(t)
 - \mathrm{i} \hat{\mathcal {L}}_{\mathrm{laser}}\hat{\rho}_{\mathrm{M}}^{(1)}(t).
\end{split}
\end{align}
The zeroth-order term $\hat{\rho}_{\mathrm{M}}^{(0)}(t)$ describes the dynamics in the absence of the laser
and based on the description above,  can be trivially obtained as
$\hat{\rho}_{\mathrm{M}}^{(0)}(t) = \hat{\rho}_{\mathrm{M,can}}$.
The first-order term $\hat{\rho}_{\mathrm{M}}^{(1)}(t)$ accounts for the dynamics induced
by the laser pulse on the incoherent superposition of eigenstates $\hat{\rho}_{\mathrm{M,can}}$.
Since
$- \mathrm{i} (\hat{\mathcal {L}}_{\mathrm{M}} - \mathrm{i}\hat{\mathcal {D}}) \hat{\rho}^{(0)}(t) =
- \mathrm{i} (\hat{\mathcal {L}}_{\mathrm{M}} - \mathrm{i}\hat{\mathcal {D}}) \hat{\rho}_{\mathrm{M,can}} =0$,
the system-bath contribution is absent here.

In deriving Eq. (\ref{equ:DentMatElemUnitEvo}), we had assumed that
$\hat{\rho}_{\mathrm{M}}(0) = |1\rangle \langle 1|$, whereas in our case we have an incoherent superposition
of eigenstates (\ref{equ:IncIniSta}).  Hence, by virtue of the linearity of the master equation
(\ref{equ:GenNonMarEvo}), we can infer directly that
\begin{align}
\label{equ:DentMatElemCanIniSta}
\begin{split}
\rho_{ab}^{(1)}(t) &= \frac{1}{Z} \frac{4 \pi}{\hbar^2} \sum_n \mathrm{e}^{-E_n \beta}
\mathrm{d}_{an} \mathrm{d}^{*}_{bn}|\epsilon(\omega_{an})| |\epsilon(\omega_{bn})|
\exp(-\mathrm{i}\omega_{ab}t)
\\ &\times
\exp\left\{\mathrm{i}\left[\phi(\omega_{an}) -\phi(\omega_{bn})
+ \frac{z}{c}(\omega_{an} - \omega_{bn}) \right] \right\}.
\end{split}
\end{align}
Therefore, it is clear that from Eq. (\ref{equ:DentMatElemCanIniSta}) that no phase information 
is encoded in the populations, $a=b$, of $\hat{\rho}_{\mathrm{M}}^{(1)}(t)$.

The second-order term $\hat{\rho}_{\mathrm{M}}^{(2)}(t)$ accounts for the evolution of the state prepared
by the laser excitation in the presence of the system-bath dynamics.
For the populations of $\hat{\rho}_{\mathrm{M}}^{(2)}$ we have
\begin{equation}
\label{equ:GenNonMarEvoDiag}
\begin{split}
\dot{\rho}_{aa}^{(2)} &=
\frac{\mathrm{i}}{\hbar} \sum_c \left(H^{\mathrm{MR}}_{ac}\rho^{(1)}_{ca} -
\rho^{(1)}_{ac} H^{\mathrm{MR}}_{ca}\right)
\\
&- \sum_{c,d}\int\limits_0^{t-t_0} \mathrm{d}\tau \left\{
\mathsf{M}_{cd,da}(-\tau)\mathrm{e}^{\mathrm{i}\omega_{da}\tau}
\rho^{(1)}_{ac}(t-\tau)
+\mathsf{M}_{ac,cd}(\tau) \mathrm{e}^{\mathrm{i}\omega_{ac}\tau}
\rho^{(1)}_{da}(t-\tau)\right.
\\
&\left. -\left[
\mathsf{M}_{da,ac}(-\tau) \mathrm{e}^{\mathrm{i}\omega_{ac}\tau}
+
\mathsf{M}_{da,ac}(\tau) \mathrm{e}^{\mathrm{i}\omega_{da}\tau}
\right]\rho^{(1)}_{cd}(t-\tau)\right\},
\end{split}
\end{equation}
Explicitly integrating Eq.~(\ref{equ:GenNonMarEvoDiag}) requires that we specify
 the system-bath interactions [Eq.~(\ref{equ:HME})] and the bath nature.
For real environments this integration is not possible and one has to appeal to specific
models;
in the Appendix we discuss the most relevant features of the commonly used bosonic
bath.
Despite the sheer complexity of integrating Eq.~(\ref{equ:GenNonMarEvoDiag}),
one can still extract some of the relevant physical contributions directly from the master
equation (as an example, see Ref.~\citenum{PB12d} for a specific implementation).
By inserting Eq.~(\ref{equ:DentMatElemCanIniSta}) into the right-hand side of
Eq.~(\ref{equ:GenNonMarEvoDiag}), \textit{it is clear that phase control of the population will result
from the system-bath interaction, which couples $\rho_{ab}^{(1)}$ to $\rho_{aa}^{(2)}$.}
This phase control is of second order in both the radiation field and  in the
system-bath coupling strength.
Note that the radiation field (the first two terms) also couples $\rho_{ab}^{(1)}$ to $\rho_{aa}^{(2)}$.
This contribution, however, is  third order in the radiation field and thus is negligible in
our current considerations of second order effects.

The time dependence of the $\mathsf{M}$ matrix elements signifies the non-Markovian character of
the dynamics, which can be associated with the environment self-correlations.
These self-correlations allows for a back and forth information flow between the system
and the bath\cite{PHEJ09}.
If the dynamics is Markovian (see next case below), the information flow is unidirectional:
from the system to the bath.   
This information flow can be understood as an entropy flux and in the case of Markovian
dynamics results in an always increasing entropy of the system M. In the
presence of external driving fields, non-Markovian terms are essential\cite{SN&11}.
In our case, this back and forth flow allows for an \emph{efficient} transfer of phase information
via the bath.
Additionally, the non-Markovian dynamics is characterized by slower correlations decay-rate
\cite{EZP12}, allowing thus the presence of phase control for longer times\cite{PB12d}.

\subsubsection{Markovian Approximation}\textemdash
If the environment quickly ``forgets" any internal self-correlations during its interaction with
the system, that is, if any dynamically established quantum correlations between parts of
the environment are destroyed on a time scale much shorter than the characteristic time
scale of the interaction-picture reduced density operator of the system, the Markovian
approximation can be invoked.
Mathematically, this means that the correlations $C_{u\nu}(\tau)$, and hence the 
$\mathsf{M}$ matrix
elements,  die off quickly as $\tau>0$, and thus the $\tau$ integral in Eq.~(\ref{equ:GenNonMarEvoDiag})
picks up contribution only from $\rho^{(1)}_{ab}(t)$, rendering the dynamics time-local.
Moreover, it does not make a difference if we extend the upper limit of integration to infinity,
in which case the matrix elements $\Gamma_{ab,cd}$, defined below, become constant.
As a result, a Markovian master equation is obtained:
\begin{equation}
\label{equ:MarEvoDiag}
\begin{split}
\dot{\rho}^{(2)}_{aa} &=
\frac{\mathrm{i}}{\hbar} \sum_c \left(H^{\mathrm{MR}}_{ac}\rho^{(1)}_{ca} -
\rho^{(1)}_{ac} H^{\mathrm{MR}}_{ca}\right)
\\
&- \sum_{c,d} \left\{
\Gamma_{ad,dc}(\omega_{cd}) \rho^{(1)}_{ac}(t)
+\Gamma_{ac,cd}(\omega_{dc}) \rho^{(1)}_{da}(t)\right.
\\
&\left. \hspace{1.5cm} -\left[
\Gamma_{ca,ad}(\omega_{da}) + \Gamma_{da,ac}(\omega_{ca})\right]
\rho^{(1)}_{cd}(t)\right\},
\end{split}
\end{equation}
where
\begin{align}
\Gamma_{ab,cd}(\omega) &= \Re \int_0^{\infty} \mathrm{d}\tau \mathrm{e}^{\mathrm{i}\omega\tau}
\mathsf{M}_{ab,cd}(\tau)
=
\Re \sum_{u,v} K_{ab}^{u} K_{cd}^{v} \int_0^{\infty} \mathrm{d}\tau
\mathrm{e}^{\mathrm{i}\omega\tau}C_{uv}(\tau).
\end{align}
This master equation retains the essential feature of the coupling of $\rho_{ab}^{(1)}$ to
$\rho_{aa}^{(2)}$ as in Eq.~(\ref{equ:GenNonMarEvoDiag}).
\textit{As a result, even given the Markovian approximation,  the population will show phase dependence of second order in the radiation
field and of second order in the system-bath coupling.}
The Markovian approximation, which results in the time locality of the master equation,
does not undermine the essential feature of phase dependence in the population.
However, Markovian dynamics results, in general, in a stronger correlations decay-rate \cite{EZP12}
and therefore, the phase information will rapidly flow back to the bath.
\textit{Hence, phase control may not be noticeable when the bath relaxation is very fast.}

\subsubsection{Markovian Secular Approximation}\textemdash
If, in addition, one now applies the secular approximation in Eq.~(\ref{equ:MarEvoDiag}), we get
\begin{align}
\begin{split}
\dot{\rho}^{(2)}_{aa} &=
\frac{\mathrm{i}}{\hbar} \sum_c \left(H^{\mathrm{MR}}_{ac}\rho^{(1)}_{ca} -
\rho^{(1)}_{ac} H^{\mathrm{MR}}_{ca}\right)
\\
&- 2
\sum_c \left[
\Gamma_{ac,ca}(\omega_{ac})\rho^{(1)}_{aa}(t) -
\Gamma_{ca,ac}(\omega_{ca})\rho^{(1)}_{cc}(t)
\right].
\end{split}
\end{align}
This approximation decouples the evolution of the diagonal elements from that of the
off-diagonal terms via the bath.  Hence no phase control in the population can be
observed, due to the artificial truncation of the off-diagonal terms.
Thus, the secular approximation is not appropriate for the study of environmentally assisted
phase control.  \textit{Alternatively, from a physics perspective, 
if the secular approximation is indeed valid for some system of interest, then 
our analysis shows that OPPC is not achievable in such a system.}

\subsection{The significant influence of the initial system-bath correlations}
\label{subsect:InfInitSysBathCorrel}
The presence of initial correlations between the system and  bath,
$\hat{\rho}_{\mathrm{ME}}(t_0) \neq \hat{\rho}_{\mathrm{M}}(t_0) \otimes \hat{\rho}_{\mathrm{E}}(t_0)$,
introduces an important new contribution in the master equation  [Eq.~(\ref{equ:GenNonMarEvo})].  For
completeness we write  the entire expression,
\begin{equation}
\label{equ:GenNonMarEvoIniCorr}
\begin{split}
\dot{\rho}_{ab} &= - \mathrm{i} \omega_{ab} \rho_{ab}
+ \frac{\mathrm{i}}{\hbar} \sum_c \left(H^{\mathrm{MR}}_{ac}\rho_{cb} -
\rho_{ac} H^{\mathrm{MR}}_{cb}\right)
\\
&- \frac{\mathrm{i}}{\hbar} \sum_u \langle a |\mathrm{tr}_\mathrm{E}
\left(\left[\hat{K}^u \hat{\Phi}^u, \hat{\rho}_{\mathrm{ME}}(t_0)\right] -
\left[\hat{K}^u \hat{\Phi}^u, \hat{\rho}_{\mathrm{E}}(t_0) \hat{\rho}_{\mathrm{M}}(t_0)\right] \right)| b \rangle
\\
&- \sum_{c,d}\int\limits_0^{t-t_0} \mathrm{d}\tau \left\{
\mathsf{M}_{cd,db}(-\tau)\mathrm{e}^{\mathrm{i}\omega_{da}\tau}
\rho_{ac}(t-\tau)
+\mathsf{M}_{ac,cd}(-\tau) \mathrm{e}^{\mathrm{i}\omega_{bc}\tau}
\rho_{db}(t-\tau)\right.
\\
&\left. -\left[
\mathsf{M}_{db,ac}(-\tau) \mathrm{e}^{\mathrm{i}\omega_{bc}\tau}
+\mathsf{M}_{db,ac}(-\tau) \mathrm{e}^{\mathrm{i}\omega_{da}\tau}
\right]\rho_{cd}(t-\tau)\right\},
\end{split}
\end{equation}
where $\hat{\rho}_{\mathrm{E}}(t_0) = \mathrm{tr}_{{\mathrm{S}}}\hat{\rho}_{\mathrm{ME}}(t_0)$
and $\hat{\rho}_{\mathrm{M}}(t_0) = \mathrm{tr}_{{\mathrm{E}}}\hat{\rho}_{\mathrm{ME}}(t_0)$.
The new term in the dynamics exposes the initial correlations between the system and
the environment.
Note that this is an extension beyond standard master equations, which do not 
usually include the state of the environment explicitly.

If $\hat{\rho}_{\mathrm{ME}}(t_0)$ factorizes into system and bath terms, then it is clear that the
terms in the second line in Eq.~(\ref{equ:GenNonMarEvoIniCorr}), the ones which account for the 
initial correlations, vanish.
In the opposite case, this would imply that the initial stationary state should be re-derived
taking into account this new driving term, i.e., the new stationary state will not be diagonal
in the energy eigenbasis.
The immediate consequence of this fact is that the initial state, before the coherent excitation
takes place, will contain stationary coherences\cite{PB12c}, i.e. non-zero time independent
off-diagonal elements of the system density matrix in the energy representation,
which will allow for phase control.

As in the previous case, we can write Eq.~(\ref{equ:GenNonMarEvoIniCorr}) as
$\dot{\hat{\rho}}_{\mathrm{M}} = -\mathrm{i} \hat{\mathcal {L}}_{\mathrm{M}}\hat{\rho}_{\mathrm{M}} 
-\mathrm{i} \hat{\mathcal {L}}_0\hat{\rho}_{\mathrm{M}} - \hat{\mathcal {D}} \hat{\rho}_{\mathrm{M}}
- \mathrm{i} \hat{\mathcal {L}}_{\mathrm{laser}}\hat{\rho}_{\mathrm{M}}$
and assume that the initial state is an eigenstate of the superoperator
$\hat{\mathcal{T}}\exp[-\mathrm{i} \int_{t_0}^{t} \mathrm{d}s(\hat{\mathcal {L}}_{\mathrm{M}}(s) + \hat{\mathcal {L}}_0(s)
- \mathrm{i}\hat{\mathcal {D}}(s))]$.
Here $\hat{\mathcal {L}}_{\mathrm{M}}$, $\hat{\mathcal {D}}$, $\hat{\mathcal {L}}_{\mathrm{laser}}$ are
defined as before and $ \hat{\mathcal {L}}_0$ denotes the contribution from the initial
correlations given in the second line of Eq.~(\ref{equ:GenNonMarEvoIniCorr}).
Due to the initial correlations between the system and the bath, it is clear that in the eigenstates
of the super operator
$\hat{\mathcal{T}}\exp[-\mathrm{i} \int_{t_0}^{t} \mathrm{d}s(\hat{\mathcal {L}}_{\mathrm{M}}(s) + \hat{\mathcal {L}}_0(s)
- \mathrm{i}\hat{\mathcal {D}}(s))]$ are no longer diagonal  in the eigenbasis of
$\hat{H}_\mathrm{M}$.
For this case,
\begin{equation}
\hat{\rho}_{\mathrm{M,eq}} = \lim_{t\rightarrow \infty}\hat{\rho}_{\mathrm{M}}(t) =
\frac{1}{Z'}\mathrm{tr}_{\mathrm{E}}\mathrm{e}^{- (\hat{H}_{\mathrm{M}} + \hat{H}^{\mathrm{ME}} + \hat{H}_{\mathrm{E}})\beta},
\end{equation}
where $Z'=\mathrm{tr}(\mathrm{e}^{-(\hat{H}_{\mathrm{M}} + \hat{H}^{\mathrm{ME}} + \hat{H}_{\mathrm{E}})\beta})$
is the partition function.
Thus, in this case the laser excitation finds  the system with stationary coherences of energy eigenstates, a concept that  we have discussed in Ref.~\citenum{PB12c}.
\textit{The presence of these off-diagonal elements guarantees that the phases of the laser pulse
contribute to the dynamics of the populations}.

Following a perturbative approach similar to that in  Sec.~\ref{subsect:InfTNonUnitDynam}, we
can study the dynamics by means of the sequence
\begin{align}
\dot{\hat{\rho}}_{\mathrm{M}}^{(0)}(t) &=
-\mathrm{i}(\hat{\mathcal {L}}_{\mathrm{M}} + \hat{\mathcal {L}}_0  - \mathrm{i}\hat{\mathcal {D}})
\hat{\rho}_{\mathrm{M,eq}},
\\
\dot{\hat{\rho}}_{\mathrm{M}}^{(1)}(t) &=
- \mathrm{i} (\hat{\mathcal {L}}_{\mathrm{M}} +  \hat{\mathcal {L}}_0 - \mathrm{i}\hat{\mathcal {D}}) \hat{\rho}_{\mathrm{M}}^{(0)}(t)
 - \mathrm{i} \hat{\mathcal {L}}_{\mathrm{laser}}\hat{\rho}_{\mathrm{M}}^{(0)}(t) ,
\\
\begin{split}
\dot{\hat{\rho}}_{\mathrm{M}}^{(2)}(t) &=
- \mathrm{i} (\hat{\mathcal {L}}_{\mathrm{M}} +  \hat{\mathcal {L}}_0 - \mathrm{i}\hat{\mathcal {D}}) \hat{\rho}_{\mathrm{M}}^{(1)}(t)
 - \mathrm{i} \hat{\mathcal {L}}_{\mathrm{laser}}\hat{\rho}_{\mathrm{M}}^{(1)}(t).
\end{split}
\end{align}
Since the off-diagonal elements in the system energy eigenbasis are present from the beginning,
 phase control of the population will take place at the time scale
of the radiation field. In other words, the phase information can
``flow" directly into the population via
the radiation, without having to wait for the system-bath interaction
to first establish phase dependence in the off-diagonal elements.

\section{Concluding Remarks}

Environmentally assisted one-photon phase control arises in open systems
when control is achieved over system properties that commute with the system
Hamiltonian. In such cases, phase control is not possible unless the
environment participates in the dynamics. 
In this paper we have analyzed the origins of such control and 
shown that the effects of molecule-bath coupling on phase
control are two-fold and  take place on two different time scales.
First, the thermalization of the coupled system-bath gives rise
to non-zero off-diagonal elements in the initial molecular state,
which can result in a direct phase control of the population; this
control takes place at the timescale of the radiation field. Second,
through the molecule-bath coupled dynamics, as modeled by Eq.~(\ref{equ:GenNonMarEvoDiag})
or Eq.~(\ref{equ:MarEvoDiag}), the phase information contained in the off-diagonals can
``flow" through the bath into the diagonal elements,
resulting in an indirect phase control; this control takes
place on  the timescale of the molecule-bath interaction. The different
timescales provide useful indicators to distinguish the two kinds
of phase control in experiments. 

One final note is in order. 
Environmentally assisted one-photon phase control depends on the system-bath interaction. 
In the case where the bath is in equilibrium, as could be the case of a solvated 
molecule, this interaction will cause, long after the pulse is over,
the system to relax to a laser-phase-independent equilibrium state as well\cite{PB12d}. 
As such, and under these circumstances, environmentally assisted one-photon phase control 
will only survive for a well defined time after the laser excitation is over. 
By contrast, if the bath comprises part of the same system as, e.g., in macromolecules,
then the induced OPPC survive for longer times assisted by the complex structure of the bath
and the strong coupling to it\cite{AB13}.
In general, rates and time scales over which this control persists is dependent upon, and requires 
numerical studies of, individual systems of interest.

\section*{Acknowledgments}
This work was supported by NSERC and by the US Air Force Office of Scientific Research under contract
number FA9550-10-1-0260, by \textit{Comit\'e para el Desarrollo de la Investigaci\'on}
--CODI-- of Universidad de Antioquia, Colombia under contract number E01651 and under the 
\textit{Estrategia de Sostenibilidad} 2013-2014 and by 
the \textit{Colombian Institute for the Science and Technology Development} --COLCIENCIAS--
under the contract number 111556934912.

\footnotesize{
\bibliography{oppcFDv2}

\providecommand*{\mcitethebibliography}{\thebibliography}
\csname @ifundefined\endcsname{endmcitethebibliography}
{\let\endmcitethebibliography\endthebibliography}{}
\begin{mcitethebibliography}{21}
\providecommand*{\natexlab}[1]{#1}
\providecommand*{\mciteSetBstSublistMode}[1]{}
\providecommand*{\mciteSetBstMaxWidthForm}[2]{}
\providecommand*{\mciteBstWouldAddEndPuncttrue}
  {\def\EndOfBibitem{\unskip.}}
\providecommand*{\mciteBstWouldAddEndPunctfalse}
  {\let\EndOfBibitem\relax}
\providecommand*{\mciteSetBstMidEndSepPunct}[3]{}
\providecommand*{\mciteSetBstSublistLabelBeginEnd}[3]{}
\providecommand*{\EndOfBibitem}{}
\mciteSetBstSublistMode{f}
\mciteSetBstMaxWidthForm{subitem}
{(\emph{\alph{mcitesubitemcount}})}
\mciteSetBstSublistLabelBeginEnd{\mcitemaxwidthsubitemform\space}
{\relax}{\relax}

\bibitem[Shapiro and Brumer(2012)]{SB11}
M.~Shapiro and P.~Brumer, \emph{Quantum Control of Molecular Processes},
  Wiley-VCH, Weinheim, 2nd edn., 2012\relax
\mciteBstWouldAddEndPuncttrue
\mciteSetBstMidEndSepPunct{\mcitedefaultmidpunct}
{\mcitedefaultendpunct}{\mcitedefaultseppunct}\relax
\EndOfBibitem
\bibitem[{Coherence and Control in Chemistry}(2011)]{faraday153}
{Coherence and Control in Chemistry}, \emph{Disc. Far. Soc.}, 2011,
  \textbf{153}, 1--428\relax
\mciteBstWouldAddEndPuncttrue
\mciteSetBstMidEndSepPunct{\mcitedefaultmidpunct}
{\mcitedefaultendpunct}{\mcitedefaultseppunct}\relax
\EndOfBibitem
\bibitem[Brumer and Shapiro(1989)]{BS89}
P.~Brumer and M.~Shapiro, \emph{Chem. Phys.}, 1989, \textbf{139},
  221--228\relax
\mciteBstWouldAddEndPuncttrue
\mciteSetBstMidEndSepPunct{\mcitedefaultmidpunct}
{\mcitedefaultendpunct}{\mcitedefaultseppunct}\relax
\EndOfBibitem
\bibitem[Prokhorenko \emph{et~al.}(2006)Prokhorenko, Nagy, Waschuk, Brown,
  Birge, and Miller]{PM06}
V.~Prokhorenko, A.~Nagy, S.~A. Waschuk, L.~S. Brown, R.~R. Birge and R.~J.~D.
  Miller, \emph{Science}, 2006, \textbf{313}, 1257--1261\relax
\mciteBstWouldAddEndPuncttrue
\mciteSetBstMidEndSepPunct{\mcitedefaultmidpunct}
{\mcitedefaultendpunct}{\mcitedefaultseppunct}\relax
\EndOfBibitem
\bibitem[Joffre(2007)]{J07}
M.~Joffre, \emph{Science}, 2007, \textbf{317}, 453b\relax
\mciteBstWouldAddEndPuncttrue
\mciteSetBstMidEndSepPunct{\mcitedefaultmidpunct}
{\mcitedefaultendpunct}{\mcitedefaultseppunct}\relax
\EndOfBibitem
\bibitem[Prokhorenko \emph{et~al.}(2007)Prokhorenko, Nagy, Waschuk, Brown,
  Birge, and Miller]{PM07}
V.~Prokhorenko, A.~Nagy, S.~A. Waschuk, L.~S. Brown, R.~R. Birge and R.~J.~D.
  Miller, \emph{Science}, 2007, \textbf{317}, 453c\relax
\mciteBstWouldAddEndPuncttrue
\mciteSetBstMidEndSepPunct{\mcitedefaultmidpunct}
{\mcitedefaultendpunct}{\mcitedefaultseppunct}\relax
\EndOfBibitem
\bibitem[Florean \emph{et~al.}(2009)Florean, Cardoza, White, Lanyi, Sension,
  and Bucksbaum]{bucksbaum}
C.~Florean, D.~Cardoza, J.~L. White, J.~K. Lanyi, R.~J. Sension and P.~H.
  Bucksbaum, \emph{Proc. Natl. Acad. Sci. U.S.A.}, 2009, \textbf{106},
  10896\relax
\mciteBstWouldAddEndPuncttrue
\mciteSetBstMidEndSepPunct{\mcitedefaultmidpunct}
{\mcitedefaultendpunct}{\mcitedefaultseppunct}\relax
\EndOfBibitem
\bibitem[Prokhorenko \emph{et~al.}(2011)Prokhorenko, Halpin, Johnson, Miller,
  and Brown]{millerreply2}
V.~I. Prokhorenko, A.~Halpin, P.~J.~M. Johnson, R.~Miller and L.~S. Brown,
  \emph{J. Chem. Phys.}, 2011, \textbf{134}, 085105\relax
\mciteBstWouldAddEndPuncttrue
\mciteSetBstMidEndSepPunct{\mcitedefaultmidpunct}
{\mcitedefaultendpunct}{\mcitedefaultseppunct}\relax
\EndOfBibitem
\bibitem[Spanner \emph{et~al.}(2010)Spanner, Arango, and Brumer]{SAB10}
M.~Spanner, C.~A. Arango and P.~Brumer, \emph{J. Chem. Phys.}, 2010,
  \textbf{133}, 151101\relax
\mciteBstWouldAddEndPuncttrue
\mciteSetBstMidEndSepPunct{\mcitedefaultmidpunct}
{\mcitedefaultendpunct}{\mcitedefaultseppunct}\relax
\EndOfBibitem
\bibitem[Grinev \emph{et~al.}(2013)Grinev, Shapiro, and Brumer]{GSB13}
T.~Grinev, M.~Shapiro and P.~Brumer, \emph{J. Chem. Phys.}, 2013, \textbf{138},
  044306\relax
\mciteBstWouldAddEndPuncttrue
\mciteSetBstMidEndSepPunct{\mcitedefaultmidpunct}
{\mcitedefaultendpunct}{\mcitedefaultseppunct}\relax
\EndOfBibitem
\bibitem[Arango and Brumer(2013)]{AB13}
C.~A. Arango and P.~Brumer, \emph{J. Chem. Phys.}, 2013, \textbf{138},
  071104\relax
\mciteBstWouldAddEndPuncttrue
\mciteSetBstMidEndSepPunct{\mcitedefaultmidpunct}
{\mcitedefaultendpunct}{\mcitedefaultseppunct}\relax
\EndOfBibitem
\bibitem[Pach\'on and Brumer(2012)]{PB12}
L.~A. Pach\'on and P.~Brumer, \emph{Phys. Chem. Chem. Phys.}, 2012,
  \textbf{14}, 10094\relax
\mciteBstWouldAddEndPuncttrue
\mciteSetBstMidEndSepPunct{\mcitedefaultmidpunct}
{\mcitedefaultendpunct}{\mcitedefaultseppunct}\relax
\EndOfBibitem
\bibitem[Pach\'on and Brumer(2011)]{PB11}
L.~A. Pach\'on and P.~Brumer, \emph{J. Phys. Chem. Lett.}, 2011, \textbf{2},
  2728\relax
\mciteBstWouldAddEndPuncttrue
\mciteSetBstMidEndSepPunct{\mcitedefaultmidpunct}
{\mcitedefaultendpunct}{\mcitedefaultseppunct}\relax
\EndOfBibitem
\bibitem[Pach\'on and Brumer(2012)]{PB12b}
L.~A. Pach\'on and P.~Brumer, \emph{J. Math. Phys. (submitted),
  arXiv:1207.3104}, 2012\relax
\mciteBstWouldAddEndPuncttrue
\mciteSetBstMidEndSepPunct{\mcitedefaultmidpunct}
{\mcitedefaultendpunct}{\mcitedefaultseppunct}\relax
\EndOfBibitem
\bibitem[Pach\'on and Brumer(2013)]{PB12c}
L.~A. Pach\'on and P.~Brumer, \emph{Phys. Rev. A}, 2013, \textbf{87},
  022106\relax
\mciteBstWouldAddEndPuncttrue
\mciteSetBstMidEndSepPunct{\mcitedefaultmidpunct}
{\mcitedefaultendpunct}{\mcitedefaultseppunct}\relax
\EndOfBibitem
\bibitem[May and K\"uhn(2001)]{MK01}
V.~May and O.~K\"uhn, \emph{Charge and Energy Transfer Dynamics in Molecular
  Systems}, Berlin: Wiley, 2001\relax
\mciteBstWouldAddEndPuncttrue
\mciteSetBstMidEndSepPunct{\mcitedefaultmidpunct}
{\mcitedefaultendpunct}{\mcitedefaultseppunct}\relax
\EndOfBibitem
\bibitem[Triana \emph{et~al.}(2012)Triana, Pach\'on, and Brumer]{PT12}
J.~F. Triana, L.~A. Pach\'on and P.~Brumer, \emph{Canonical Typicality
  Deviations at Low Temperature}, 2012,  In preparation.\relax
\mciteBstWouldAddEndPunctfalse
\mciteSetBstMidEndSepPunct{\mcitedefaultmidpunct}
{}{\mcitedefaultseppunct}\relax
\EndOfBibitem
\bibitem[Pach\'on and Brumer(2013)]{PB12d}
L.~A. Pach\'on and P.~Brumer, \emph{J. Chem. Phys. (submitted),
  arXiv:1308.1843}, 2013\relax
\mciteBstWouldAddEndPuncttrue
\mciteSetBstMidEndSepPunct{\mcitedefaultmidpunct}
{\mcitedefaultendpunct}{\mcitedefaultseppunct}\relax
\EndOfBibitem
\bibitem[Breuer \emph{et~al.}(2009)Breuer, Laine, and Piilo]{PHEJ09}
H.-P. Breuer, E.-M. Laine and J.~Piilo, \emph{Phys. Rev. Lett.}, 2009,
  \textbf{103}, 210401\relax
\mciteBstWouldAddEndPuncttrue
\mciteSetBstMidEndSepPunct{\mcitedefaultmidpunct}
{\mcitedefaultendpunct}{\mcitedefaultseppunct}\relax
\EndOfBibitem
\bibitem[Schmidt \emph{et~al.}(2011)Schmidt, Negretti, Ankerhold, Calarco, and
  Stockburger]{SN&11}
R.~Schmidt, A.~Negretti, J.~Ankerhold, T.~Calarco and J.~T. Stockburger,
  \emph{Phys. Rev. Lett.}, 2011, \textbf{107}, 130404\relax
\mciteBstWouldAddEndPuncttrue
\mciteSetBstMidEndSepPunct{\mcitedefaultmidpunct}
{\mcitedefaultendpunct}{\mcitedefaultseppunct}\relax
\EndOfBibitem
\bibitem[Estrada \emph{et~al.}(2013)Estrada, Zueco, and Pach\'on]{EZP12}
A.~F. Estrada, D.~Zueco and L.~A. Pach\'on, \emph{Out-of-Equilibrium
  Non-Markovian Quantum Limit}, 2013,  In preparation.\relax
\mciteBstWouldAddEndPunctfalse
\mciteSetBstMidEndSepPunct{\mcitedefaultmidpunct}
{}{\mcitedefaultseppunct}\relax
\EndOfBibitem
\end{mcitethebibliography}
\bibliographystyle{rsc} 
}

\section*{Appendix: Harmonic Oscillator Bath}
Since the explicit calculation of the correlation functions $C_{u\nu}$ in Eq.~(\ref{equ:Cunu})
is not feasible for real baths such as solvents, one has to appeal to specific models of the
bath and the system-bath interaction.
The most common approximation is the normal mode approximation.
In this case one assumes that, e.g., the vibrational modes can be described by small oscillations
around the equilibrium point, thus allowing the use of the harmonic approximation.
In this case, we replace the interaction term $H^{\mathrm{ME}}$ in Eq.~(\ref{equ:HME}) by
\begin{equation}
\hat{H}^{\mathrm{ME}} = \hat{K}(s)  \hat{\Phi}(\mathbf{Z}) =
\hat{K}(s) \sum_{\xi} \hbar \,\gamma_{\xi}\,\hat{Z}_{\xi},
\end{equation}
where $s$ denotes the system coordinates, $\gamma_{\xi}$ denote the system-environment
coupling constants and $\hat{\mathbf{Z}}=\{\hat{Z}_{\xi}\}$ denotes the environment normal
mode coordinates.
For simplicity, we have assumed that $H^{\mathrm{ME}}$ contains a single term and
have dropped the index $u$.
Note that since $\langle Z_{\xi} \rangle_{\mathrm{R}} = 0$, the expression for $C_{u\nu}$ in
Eq.~(\ref{equ:Cunu}) reduces to the two-point correlation function of the bath operator.

Thus, we have for $C_{u\nu}(t) = C(t)$
\begin{equation}
C(t) = \int_0^{\infty} \mathrm{d}\omega\left[\cos(\omega t)
\coth\left(\frac{\hbar \omega}{2k_{\mathrm{B}}T} \right)
- \mathrm{i} \sin(\omega t) \right] \omega^2 J(\omega),
\end{equation}
where $J(\omega)$ is the spectral density,
\begin{equation}
J(\omega) = \sum_{\xi} g_{\xi}^2 \delta(\omega_{\xi} - \omega).
\end{equation}
Some typical spectral densities are, cf. Ref.~\citenum{PB12},
$\omega^2 J(\omega) = \theta(\omega) j_0 \omega^p \mathrm{e}^{-\omega/\omega_{\mathrm{c}}}$
or
$\omega^2 J(\omega) = \theta(\omega) \frac{j_0 \omega}{\omega^2 + \omega_{\mathrm{D}}}$,
 where $\omega_{\mathrm{c}}$ and  $\omega_{\mathrm{D}}$ both denote a cut off frequency.

\end{document}